\documentstyle[pre,aps,epsf,amssymb,preprint]{revtex}
\begin{document}
\preprint{TIFR/TH/00--62}
\draft

\title{Distribution of sizes of erased loops of loop-erased random walks in\\
two and three dimensions}

\author{Himanshu Agrawal\cite{HA} and
	Deepak Dhar\cite{DD}}
\address{Theoretical Physics Group, Tata Institute of Fundamental
	 Research, Homi Bhabha Road, Mumbai -- 400 005, India}
\date{\today}

\maketitle
\begin{abstract}

We show that in the loop-erased random walk problem, the exponent characterizing
probability distribution of areas of erased loops is superuniversal. In
$d$-dimensions, the probability that the erased loop has an area $A$ varies as
$A^{-2}$ for large $A$, independent of $d$, for $2 \le d \le 4$. We estimate the
exponents characterizing the distribution of perimeters and areas of erased
loops in $d = 2$ and $3$ by large-scale Monte Carlo simulations. Our estimate of
the fractal dimension $z$ in two-dimensions is consistent with the known exact
value $5/4$. In three-dimensions, we get $z = 1.6183 \pm 0.0004$. The exponent
for the distribution of durations of avalanche in the three-dimensional abelian
sandpile model is determined from this by using scaling relations.

\end{abstract}
\pacs{64.60.Ak, 05.20.-y, 05.40.+j, 75.10.Hk}

\narrowtext

\section{Introduction}\label{Sec:Intro}

The loop erased random walk (LERW) problem was first defined by Lawler
\cite{LAW80} as a more tractable variant of the well-known self-avoiding walk
problem. The problem turns out to be related to many well-studied problems in
statistical physics, but seems to have attracted less attention than it
deserves. It was shown by Lawler \cite{LAW87} to be equivalent to a special case
of the Laplacian self-avoiding walk problem defined by Lyklema {\em et~al.\/}
\cite{LYK86}. Majumdar \cite{MAJ92} showed that this model is equivalent to the
classical graph-theoretical problem of spanning trees on graphs, and the
$q$-state Potts model in the limit $q \to 0$. This equivalence also makes this
problem related to the abelian sandpile model of self-organized criticality
\cite{DD99}. In fact, as we shall show in this paper, this model provides a
numerically efficient method of determining the only unknown critical exponent
of the abelian sandpile model in three-dimensions.

This prompted us to undertake the numerical study of the LERW's in $d = 2$ and
$3$ reported in this paper. We obtain fairly precise estimates of the fractal
dimension of LERWs in $d = 2$ and $3$. We note the interesting consequence of
the scaling theory that the distribution of the area of the erased loop has the
same exponent $2$, {\em independent of the dimension\/} $d$, for $2 \le d \le
4$. In $d = 3$, the numerical value of fractal dimension of LERW's enables us to
determine the avalanche durations exponent of the abelian sandpile model, using
the scaling relations and other exactly known exponents of the model
\cite{KTIT00}.

A good review of earlier results on the LERW problem can be found in
\cite{LAWbook}. Lawler showed that the fractal dimension $z$ of LERWs is $2$ for
$d \ge 4$, and $z \le (d+2)/3$ for $d \le 4$ \cite{LAWbook}. Recently, it was
shown rigorously that in two-dimensions $z$ is strictly larger than 1
\cite{Law98}. Using the known exact results about the critical exponents of
Potts model from conformal field theory, Majumdar was able to prove exactly that
$z = 5/4$ for LERW problem in $d = 2$ \cite{MAJ92}, a result which was guessed
earlier by Guttmann and Bursill from numerical simulations \cite{GUT90}. A proof
of this result without using conformal field theory has been given by Kenyon
\cite{kenyon}. Using conformal invariance, Duplantier has obtained the exact
probabilities of no intersection of $n$ LERWs of $\ell$ steps starting near each
other in two-dimensions, and also the winding angle distribution \cite{DUP92}.
The distribution of sizes of erased loops was first studied in \cite{DD97}.
Priezzhev has used bounds on intersection probability of loop erased walks with
random walks to show that the upper-critical dimension of the
Bak-Tang-Wiesenfeld (BTW) sandpile model is 4 \cite{PR99}.

The plan of this paper is as follows: The LERW model is defined in Sec.\
\ref{Sec:Model}. In Sec.\ \ref{Sec:Scale}, we recall the main points of the
scaling theory of the distribution of erased loops \cite{DD97}, and apply it to
show that the exponent characterizing the distribution of area enclosed by
erased loop is the same for $2 \leq d \leq 4$. We determine the behavior of the
distribution functions for the perimeter and the area of the loop in the scaling
limit, for very small or very large values of the argument of the scaling
functions. The simulation technique and the results obtained are described in
Sec.\ \ref{Sec:Sim}. The exponent characterizing the distribution of durations
of avalanches in BTW sandpile model in $d = 3$ is determined in Sec.\
\ref{Sec:Sand}, and some concluding remarks follow in Sec.\ \ref{Sec:Conc}.

\section{Definition of the Model} \label{Sec:Model}

\begin{figure}
\centerline{\epsfxsize0.6\hsize\epsfbox{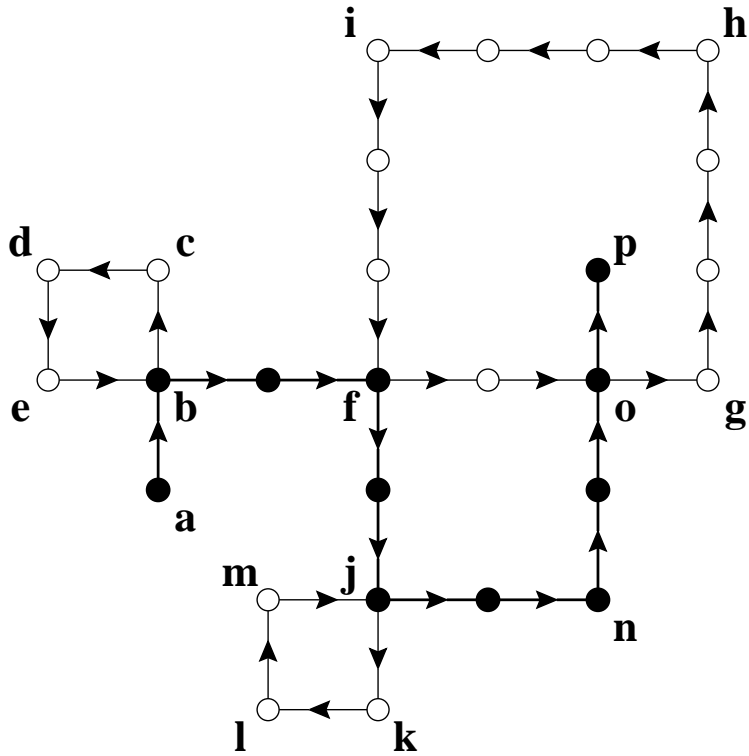}}
\vskip1ex

\caption{An illustrative example of the loop-erasing procedure: The random walk
{\bf a-b-c-d-e-$\cdots$-p} starts at {\bf a}, and ends at {\bf p}. The erased
loops are shown by thin lines. Note that at the point {\bf o}, while the random
walk path intersects itself, the LERW encounters no intersection, as the loop
{\bf f-g-h-i-f} has already been erased.} 

\label{F:LoopEr}
\end{figure}

Consider a simple random walk on a $d$-dimensional lattice. We start with a
particular realization ${\cal W}$ of the random walk having $N$ steps, ${\cal W}
= \lbrace w_{0}$, $w_{1}$, $w_{2}$, $\ldots$, $w_{N-1}$, $w_{N}\rbrace$, where
$w_{i}$ is the site reached by the $i$-th step of walk. We define the LERW
${\cal L}$ corresponding to ${\cal W}$ as the path obtained from ${\cal W}$ by
erasing each loop as soon as it is formed. If ${\cal W}$ has no
self-intersections, we define ${\cal L} = {\cal W}$. If ${\cal W}$ has
self-intersections, let $j$ be earliest step which leads to self-intersection in
${\cal W}$, so that $j$ is the least integer such that $w_{j} = w_{i}$ for some
$i < j$. Then, we obtain a new walk ${\cal W}' = \lbrace w_{0}$, $w_{1}$,
$\ldots$, $w_{i}$, $w_{j-1}$, $\ldots$, $w_{N-1}$, $w_{N}\rbrace$ by deleting
all steps between $i$ and $j$, keeping $i$ and deleting $j$. This process,
corresponding to loop erasure of the earliest loop formed, is repeated till
loops can no longer be found. The resulting walk ${\cal L}$ is the required LERW
corresponding to ${\cal W}$. This procedure of loop-erasure is illustrated in
Fig.~\ref{F:LoopEr}.

The length of ${\cal L}$ is the number of steps in ${\cal L}$. We will denote it
by $n$. For a fixed $N$, $n$ is a random variable. We define the critical
exponent $z$ of the LERW by the relation that
\begin{equation} \label{E:nNz}
  \langle n\rangle \sim N^{z/2}
\end{equation}
for large $N$, where the angular brackets denote averaging over all random walks
of $N$ steps. As the root-mean-square end to end distance $R$ is same as for
random walks, we have $R \sim N^{1/2}$, and $ \langle n \rangle \sim R^z$. Thus,
$z$ is the fractal dimension of the LERW.

\section{Scaling Theory for the Distribution of Loop-sizes}
\label{Sec:Scale}

Let $\text{Prob}(\ell,N)$ denote the probability that a loop of perimeter $\ell$
will be erased at the $N$-th step of the random walk. Let $F(\ell,N)$ denote the
cumulative probability that a loop of perimeter $\ell$ {\em or greater\/} will
be erased at the $N$-th step of the random walk. We shall study the behavior of
this function for large $N$, and write
\begin{equation}
  \text{Prob}(\ell) = \lim_{N \to \infty} \text{Prob}(\ell,N)
\end{equation}
and
\begin{equation}
  F(\ell) = \lim_{N \to \infty} F(\ell,N)
\end{equation}

We adopt the convention that if no loop is formed at a step, it will be said to
be erasure of loop of perimeter $0$. With this convention, we clearly have
$F(0,N) = 1$, for all $N$.

For $d \le 4$, the mean number of loop-length erased per step tends to $1$ for
large $N$. This implies that
\begin{equation}
  \sum_{\ell = 0}^{\infty} \ell \text{~Prob}(\ell) = 1, \quad \text{for~}
d \le 4
\end{equation}

For large $\ell$, $F(\ell)$ is expected to vary as a power of $\ell$, say as
$\ell^{-\tau +1}$. However, for a finite $N$, there is a cutoff size
$\ell^{\star}$, and loops of size $\ell > \ell^{\star}$ are very unlikely. The
cutoff value $\ell^{\star}$ varies as a power of $N$, say $\ell^{\star} \sim
N^{\phi}$. This suggests that $F(\ell,N)$ satisfies the scaling form
\begin{equation}\label{E:FlN_FlNPhi}
  F(\ell,N) \sim \ell^{-\tau +1} f\bigl(\ell/N^{\phi}\bigr)
\end{equation}

The cutoff exponent $\phi$ can be determined by the following simple argument
\cite{DD97}: The cutoff for the perimeter of erased loops should also vary as
$\langle n\rangle$, the average length of the LERW after $N$ steps. Since this
scales as $N^{z/2}$ [Eq.~(\ref{E:nNz})], we get $\phi = z/2$.

The exponent $\tau$ is also expressible in terms of $z$. For $\ell \ll
\ell^{\star}$, the total number of loops of size $ \geq \ell$ for a walk of $N$
steps varies as $N F(\ell)$, and is much greater than $1$. For $\ell >
\ell^{\star}$, we expect a much stronger decay. For $2 \le d \le 4$, we get a
significant number of large loops, and thus in this case, we expect that
\begin{equation}
  N F(\ell^{\star},N) \sim O(1)
\end{equation}
Putting in the scaling form (\ref{E:FlN_FlNPhi}), this implies that
\begin{equation}
  \tau = 1 + 2/z
\end{equation}

Thus, the scaling form for the distribution of loop perimeters is determined in
terms of a single exponent $z$, and is given by 
\begin{equation}\label{E:FlN}
  F(\ell,N) \sim \ell^{-2/z} f\bigl(\ell/N^{z/2}\bigr),
            \quad \text{for~} \ell \gg 1.
\label{scaling_eq}
\end{equation}

The scaling function $f(x)$ is assumed to tend to $1$ as $x$ tends to zero, and
tend to zero for large $x$. We define 
\begin{equation}
  \Delta \text{Prob}(\ell,N) = \text{Prob}(\ell,N) - \text{Prob}(\ell)
\end{equation}

If for $x$ near zero, $1-f(x)$ varies as $ x^a$, we see that keeping $\ell$
fixed, and in the limit of large $N$
\begin{equation}
  \Delta\text{Prob}(\ell,N) \sim - K_{\ell} N^{-az/2}
\label{f_near_zero}
\end{equation}
where $K_{\ell}$ is an $\ell$-dependent constant, and the exponent is
independent of $\ell$. It is easy to calculate $\text{Prob}(2,N)$ in arbitrary
dimension $d$. The conditional probability of forming a loop of perimeter $2$ at
the $N$-th step is $0$, if the random walker returned to origin at step $(N-1)$,
and it is $1/(2d)$ otherwise (for a $d$-dimensional hypercubical lattice with
coordination number $2d$). Thus
\begin{equation}
  \text{Prob}(2,N) = \frac{1}{2d} (1 - g_{N})
\end{equation}
where $g_{N}$ is the probability that the random walker returns to origin after
$N-1$ steps. In $d$-dimensions, $g_{N}$ varies as $N^{-d/2}$ for large $N$.
Thus, we see that for large $N$, $\Delta \text{Prob}(2,N)$ varies as $N^{-d/2}$.
Comparing this with Eq.~(\ref{f_near_zero}), we see that $a = d/z$. Thus, we get
\begin{equation}
  f(x) \simeq 1 - K x^{d/z}, \quad \text{for~} x \text{~near~} 0.
\label{E:fx}
\end{equation}
[We shall denote an undetermined constant by $K$. Its value in different
equations need not be the same.] For other values of $\ell \neq 2$, this then
implies that
\begin{equation}
  K_{\ell} \simeq K \ell^{d/z}, \quad \text{for~} \ell \gg 1.
\end{equation}
This may be understood as follows: The main deviation of $\text{Prob}(\ell,N)$
from its asymptotic value comes from the cases when the LERW at step $N-1$ is of
length $\lesssim \ell$, and the probability that walker after $N$ steps is
within a sphere of radius $\ell^{1/z}$ centered at the origin varies as
$\ell^{d/z} N^{-d/2}$ for $\ell \ll N$.

We can also determine the leading $N$-dependence of $\text{Prob}(\ell=0,N)$.
Since for any non-zero $\ell$, $\text{Prob}(\ell,N)$ is less than than its
limiting value for large $N$, $\text{Prob}(0,N)$ must be larger than
$\text{Prob}(0)$. In fact
\begin{eqnarray}
  \Delta \text{Prob}(0,N)
    & =    & -\sum_{\ell=2}^{\infty} \Delta \text{Prob}(\ell,N)
\nonumber\\
    & \sim & K N^{-d/2} \sum_{\ell} \ell^{(d-2 -z)/z}
\end{eqnarray}
This summation over $\ell$ has an upper cutoff proportional to $\ell^{\star}$.
In two dimensions, we get
\begin{equation}
  \Delta \text{Prob}(\ell=0,N) \sim K(\log N)/N
\end{equation}
In three dimensions, the summation diverges as $(\ell^{\star})^{1/z} \sim
N^{1/2}$. Thus, we get
\begin{equation}
  \Delta \text{Prob}(\ell = 0,N) \sim K/N, \quad \text{for~} d=3.
\end{equation}

For large $x$, $f(x)$ is expected to decrease as $\exp(-K x^b)$. The exponent
$b$ can be determined as follows: We note that for any constant $\epsilon \ll
1$, the probability that a loop of perimeter $\epsilon N$ is formed at $N$-th
step should vary as $\exp\{-K(\epsilon) N\}$ for fixed $\epsilon$ and $N$
tending to infinity \cite{fisher}. This implies that $b = 2/(2 - z)$, and hence
\begin{equation}
  f(x) \sim \exp\Bigl(-K x^{2/(2 - z)}\Bigr), \quad \text{for large~} x.
\end{equation}

An interesting quantity is the area enclosed by a loop. In two-dimensions, this
is straightforward to determine. In three-dimensions, it may be defined as the
minimum number of plaquettes required to form a simply-connected surface bounded
by the loop. In this study, we used an alternate, computationally simpler,
measure of this area. We simply project the loop on to the three coordinate
planes, and measure the areas of the projections. If the three areas are
$a_{1}$, $a_{2}$ and $a_{3}$, we define the area of the loop to be $(a_{1}^{2} +
a_{2}^{2} + a_{3}^{2})^{1/2}$. The generalization to higher dimensions is
obvious.

Let $F(A,N)$ be the probability that a loop of area greater than or equal to $A$
generated at the $N$-th step of the random walk. A loop of perimeter $\ell$ has
a linear size $R \sim \ell^{1/z}$ and an area $A \sim R^2$, then, it is easy to
see from Eq.~(\ref{E:FlN}) that for $N, A \gg 1$
\begin{equation}
  F(A,N) \sim A^{-1} g(A/N)
\end{equation}
Here also the scaling function $g(x)$ goes to a constant for $x\to0$, and
decreases rapidly to zero for $x \gg 1$. 

Thus we find the rather unexpected result that the distribution for the area of
the loop is independent even of $z$, and hence is the same for all dimensions
$d$, with $2 \le d \le 4$. This argument does not work in $d = 1$, as there
$\text{Prob}(\ell)$ decreases exponentially with $\ell$, and the scaling theory
assuming power-law decays fails \cite{footnote}. For $d \geq 4$, the LERW
behaves as a random walk, and for random walks, the area of loop varies as the
perimeter of the loop. Hence we would expect that the probability that a loop of
area $A$ is formed varies as $A^{-d/2}$ for $d > 4$. The probability that a loop
of area greater than or equal to $A$ is formed, varies as $A^{-d/2 + 1}$ for $d
> 4$.

Using the fact that $A$ varies as $\ell^{2/z}$ for $\ell \lesssim \ell^{\star}$,
from Eq.~(\ref{E:fx}) we see that the function $g(x)$ determining the finite-$N$
cutoff effects varies as
\begin{equation}\label{E:gsmall}
g(x) \simeq g(0) \exp\bigl(-K x^{d/2}\bigr), \quad \text{for small~} x.
\end{equation}

For large $x$, $g(x)$ should vary as $\exp(-K x^{c})$, where $c$ is some
exponent. Let $\epsilon$ be a small number $\ll 1$. Using the fact that loops of
area $\epsilon L^{2}$ should decrease only as $\exp\{-K(\epsilon) L\}$ for fixed
$\epsilon$ in the limit of large $L$, we see that $c = 1$, and
\begin{equation}\label{E:glarge}
  g(x) \sim \exp(-Kx), \quad \text{for large~} x.
\end{equation}
It is interesting to compare this behavior with that of $f(x)$ for large $x$.
These behaviors are consistent only if for a large loop of area $A \gg N$,  the
average perimeter varies as 
\begin{equation}
  \ell \sim \ell^{\star} (A/N)^{1- z/2}
\end{equation}

This behavior should be contrasted with the behavior for $A \ll N$, where the
average perimeter varies as $A^{z/2}$ with no explicit dependence on
$\ell^{\star}$. It is interesting to note that this scaling law for large loops
remains valid even outside the scaling limit for $A$ of order $N^\alpha$ with $1
\leq \alpha \leq 2$. For $\alpha = 2$, it gives perimeter proportional to $N$,
as it should.

\section{Numerical Simulations} \label{Sec:Sim}

The simplest algorithm to simulate the LERW problem on computer is to actually
generate the trail of a random walk step-by-step on a $d$-dimensional lattice.
At each new step, if a loop is formed it is erased. This is straight-forward to
implement, but requires large memory in large dimensions $d$, as for simulating
a walk of $N$ steps, one needs to have a lattice of linear size $N^{1/2}$, which
means that the required memory increases as ${\cal O}\bigl(N^{d/2}\bigr)$.

The algorithm that makes the most efficient use of memory would store the walk
as a linked list, keeping only the unerased steps. But there is a memory/CPU
tradeoff, and the computation time increases as searching for self-intersections
is very inefficient in this scheme.

In our simulations we used a hybrid scheme for storing the coordinates of the
points visited by the walk. We store the coordinates of the LERW in not one, but
$M$ lists, where $M$ is a large number. There is a unique hashing rule which
assigns a site to one of the lists, so that checking for intersection has to be
done only within one list. To see if a point already belongs to the LERW, we
have to search only in the list corresponding to the point in question. The best
choice of $M$ is ${\cal O}(\ell^{\star})$, as then each list has ${\cal O}(1)$
entries. With this, we were able to simulate a two-dimensional walk of $2^{33}$
steps in about 3 hours 16 minutes using $\sim 60$ Mb of memory on a 350 MHz
Pentium-II machine. In three-dimensions, a walk of $2^{29}$ steps took about
$24$ minutes and $\sim 300$ Mb of memory on a similar machine.

Simulations were carried out for total walk length $N$ of $2^{r}$ steps, with $r
= 25, \ldots, 29$ for two-dimensional walks and $r = 24, \ldots, 28$ for
three-dimensional walks. To eliminate the initial transients, we collected the
statistics of loops only after discarding the first $N/2$ steps. In addition,
ensemble average was taken over $10^3$ distinct realizations of random walks in
each case. For the two-dimensional case we also simulated a small number of
walks for $r$ up to $34$.

\subsection{Two-dimensional LERW}

\begin{figure}
\centerline{\epsfxsize\hsize\epsfbox{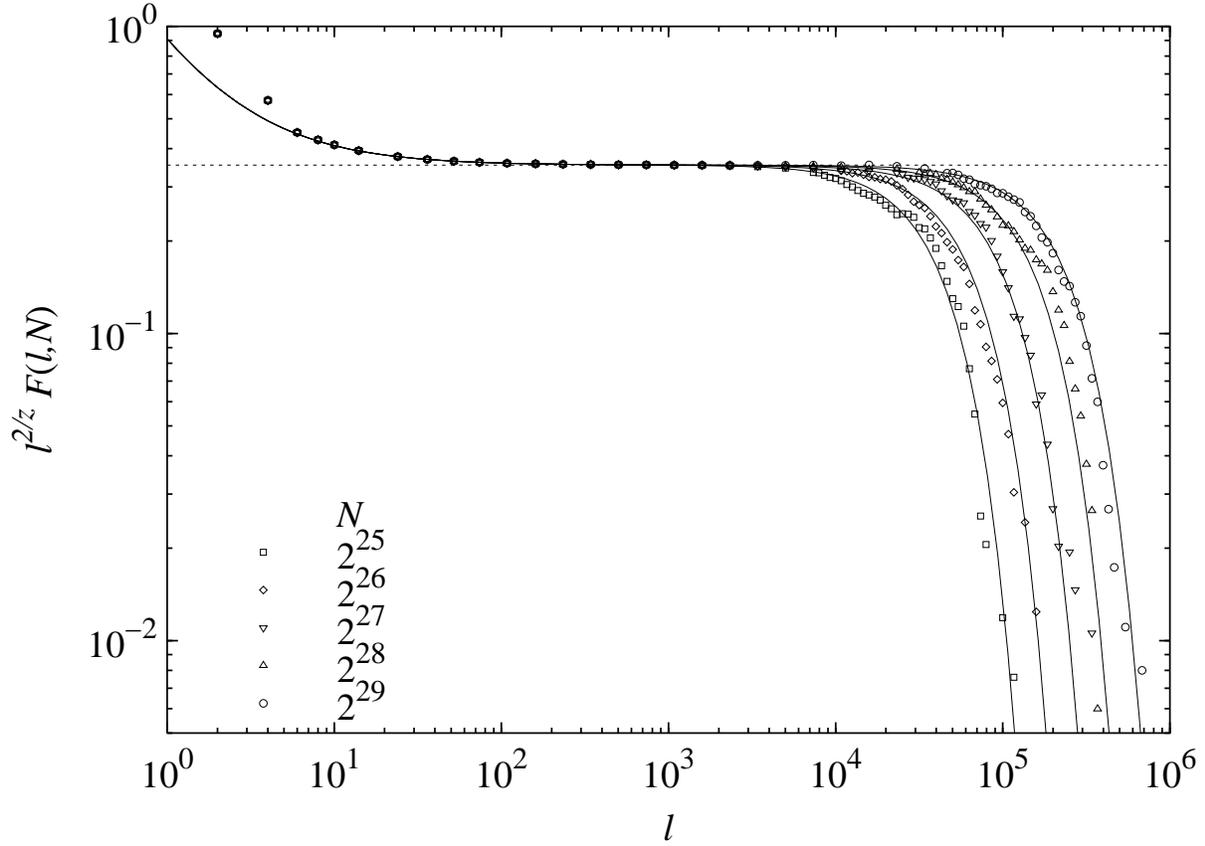}}
\vskip1ex

\caption{Variation of $\ell^{2/z}F(\ell,N)$ with $\ell$ for two-dimensional
LERW. Solid lines represent the best fit of Eq.~(\ref{E:F0}) with parameter
values (including estimate of $z$) given in Table~\ref{T:EstPar}.}

\label{F:Fl2d}
\end{figure}

In Fig.~\ref{F:Fl2d} we show the plot for the cumulative distribution function
$F(\ell,N)$ \cite{note1}. We plot $\ell^{2/z}F(\ell,N)$ versus $\ell$. There is
a significant deviation from simple power-law behavior for very small $\ell$,
and for large $\ell$. For $\ell \lesssim \ell^{\star}$, the data fits well to
the functional form given by Eq.~(\ref{E:fx}). In the small $\ell$ regime, the
leading correction is a correction to scaling. Incorporating this, we fit the
data to the form
\begin{equation}
F_{0}(\ell,N)
  = \frac{C_{1}}{\ell^{C_{2}}}
    \exp\left[
          -C_{3} \bigl(\ell^{C_{2}}/N\bigr)^{-d/2}
        \right]
    \left(
      1+\frac{C_{4}}{\ell^{C_{5}}}
    \right)
\label{E:F0}
\end{equation}
where $C_{2}$ is related to the fractal dimension via $C_{2} = 2/z$.

The best fit values of all the parameters in Eq.~(\ref{E:F0}) are tabulated in
Table~\ref{T:EstPar}. We note that $C_{5}$ is $1$ within our error bars.
Furthermore,  the exact value of $C_{2}$ is also known to be $8/5$. As a result,
one more set of values were estimated for the parameters by constraining $C_{2}$
and $C_{5}$ to these values. The parameter values thus obtained are also
tabulated in Table~\ref{T:EstPar}. The fit is rather good for all $\ell \gtrsim
10$. Statistical fluctuations are large for $\ell \gtrsim 10^4$, as there are
not many such loops generated.

\begin{figure}
\centerline{\epsfxsize\hsize\epsfbox{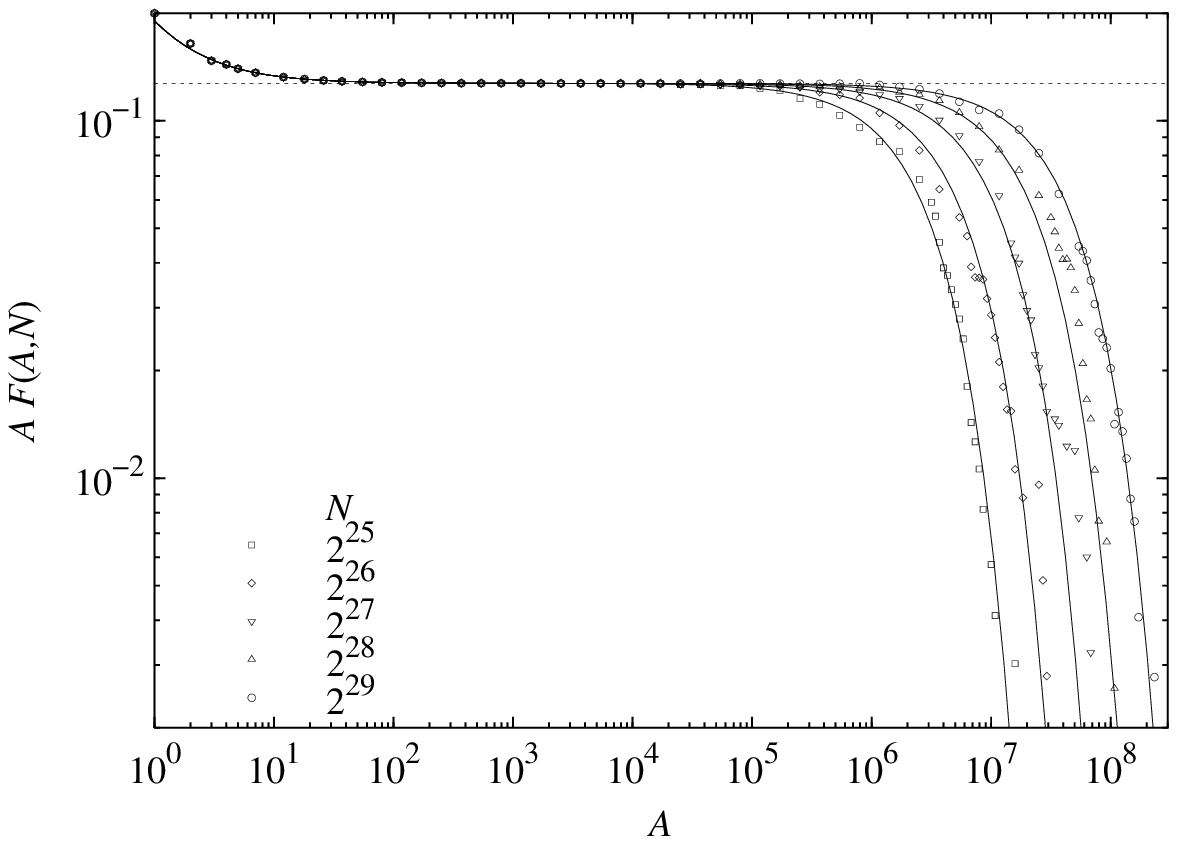}}
\vskip1ex

\caption{Variation of $A F(A,N)$ with $A$ for two-dimensional LERW. Solid lines
represent the best fit of Eq.~(\ref{E:F0}) with parameter values given in
Table~\ref{T:EstPar}.}

\label{F:FA2d}
\end{figure}

In Fig.~\ref{F:FA2d} we have plotted the $A F(A,N)$ versus $A$ for different
values of $N$, and also shown the best fit using the fitting form
Eq.~(\ref{E:F0}) with $\ell$ substituted by $A$. While estimating the parameters
we constrained $C_{2}$ and $C_{5}$ to $1$. This is because the exact value of
$C_{2}$ is known to be $1$ and unconstrained value of $C_{5}$ turns out to be
$1$ within error bars. This allows a better estimate of the remaining
parameters. The estimated best fit values of parameters for this data set are
tabulated in Table~\ref{T:EstPar}. It is clearly seen from Fig.~\ref{F:FA2d}
that the scaling form fits the data very well in nearly the entire range.

\begin{figure}
\centerline{\epsfxsize\hsize\epsfbox{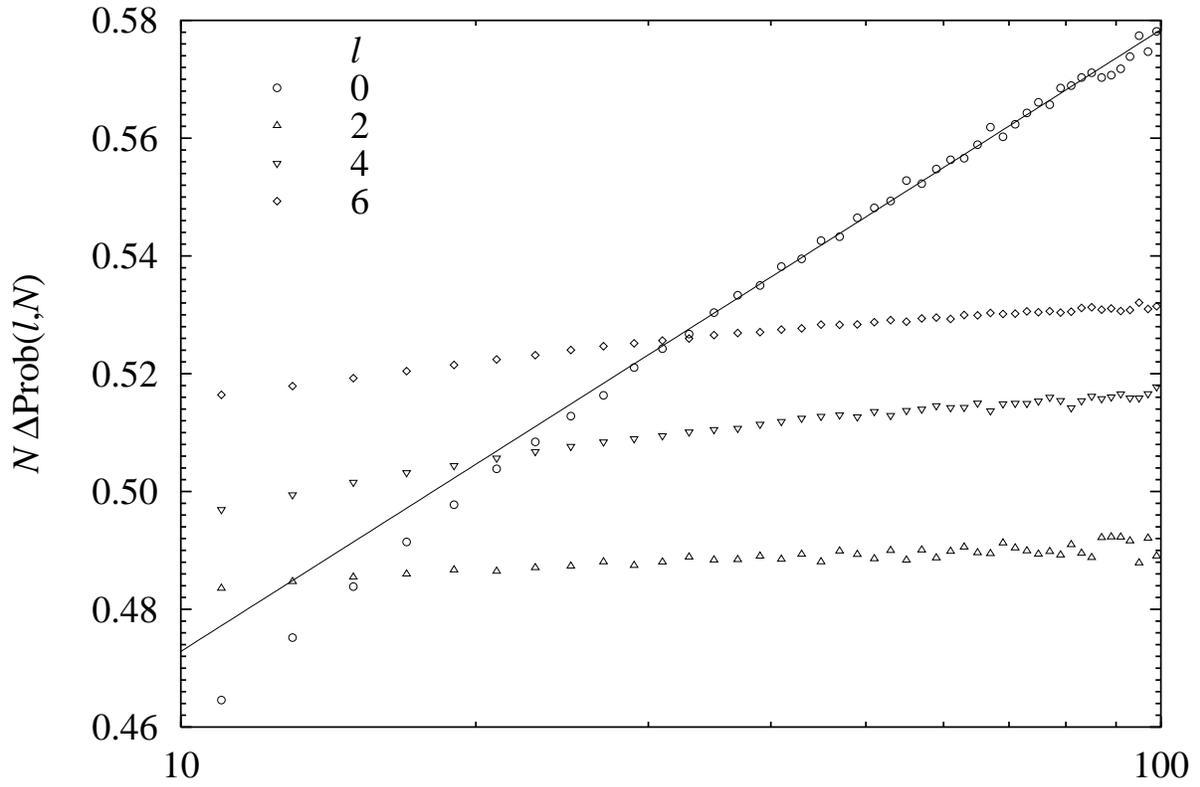}}

\caption{Variation of $N \Delta \text{Prob}(\ell,N)$ with $N$ for $\ell= 0$,
$2$, $4$ and $6$ in two dimensions. For better visibility, the data points for
$\ell=2$, $4$, and $6$ have been displaced vertically by $0.69$, $0.59$, and
$0.57$ respectively.}

\label{F:dP2d}
\end{figure}

We obtained more accurate estimates of $\text{Prob}(\ell,N)$ for $\ell \leq 100$
by taking an ensemble average over $10^{9}$ different realizations of the random
walk. In Fig.~\ref{F:dP2d}, we have plotted the variation of $N \Delta
\text{Prob}(\ell,N)$ versus $\log(N)$ in two dimensions for $\ell = 0$, $2$, $4$
and $6$. We see clearly that while $N \Delta \text{Prob}(0,N)$ has a linear
variation with $\log(N)$, for other values of $\ell$, this tends to a limiting
constant value for $\ell \gg 1$.

\subsection{Three-dimensional LERW}

\begin{figure}
\centerline{\epsfxsize\hsize\epsfbox{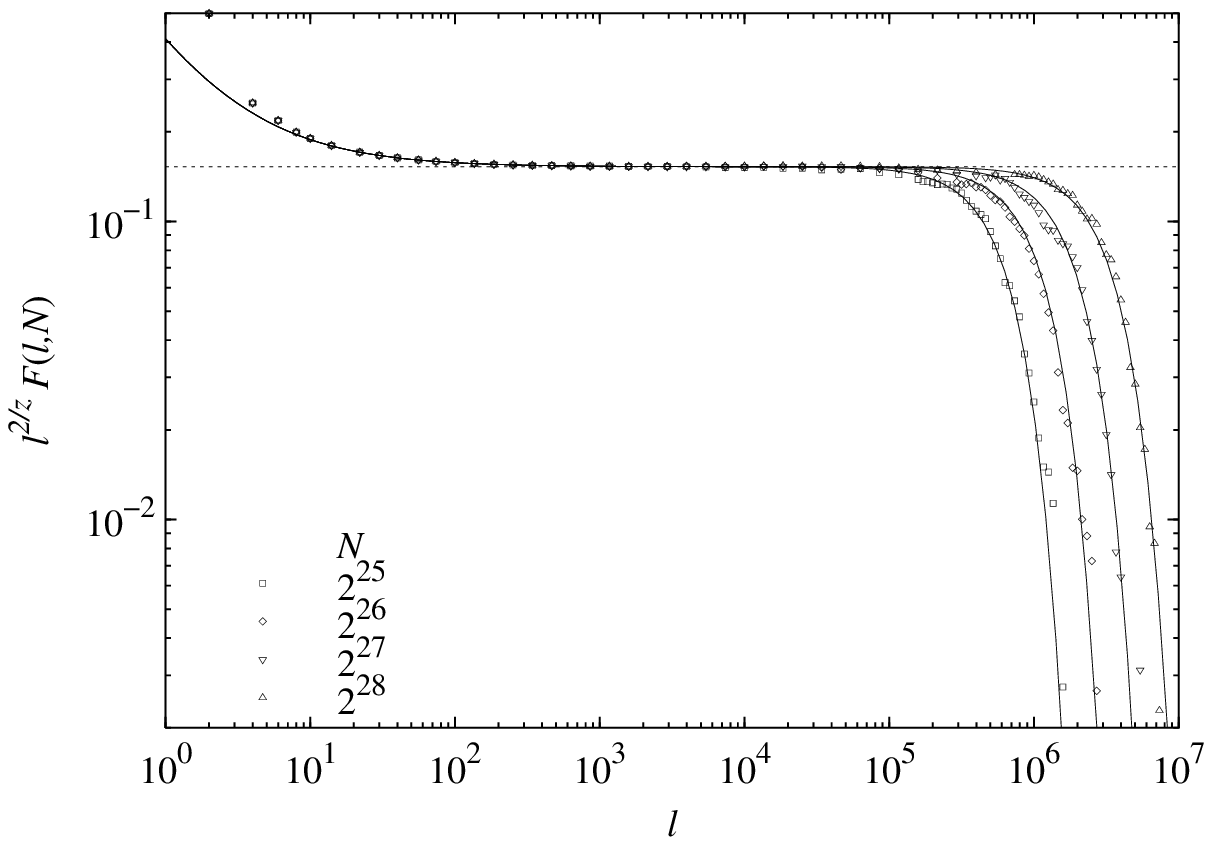}}
\vskip1ex

\caption{Variation of $\ell^{2/z}F(\ell,N)$ with $\ell$ for three-dimen\-sional
LERW. Solid lines represent the best fit of Eq.~(\ref{E:F0}) with parameter
values given in Table~\ref{T:EstPar}.}

\label{F:Fl3d}
\end{figure}

The distribution of loop-sizes for the three-dimensional walks by perimeter is
shown in Fig. \ref{F:Fl3d}. The format of presentation is exactly the same as in
the previous subsection. We fit the data to the form given by Eq.~(\ref{E:F0}).
From the figure it is seen that this scaling form fits the entire data very well
for $\ell \gtrsim 10$. The best fit values of parameters in this equation are
tabulated in Table~\ref{T:EstPar}. We find that in this case the best-fit value
of the correction to scaling exponent $C_{5}$ turns out to be $0.86$, clearly
different from $1$.

Our estimate of the best fit value of the fractal dimension $z$ gives
\begin{equation}
  z = 1.6183 \pm 0.0004, \qquad d = 3
\end{equation}
This value is not very sensitive to the choice of parameters $C_{1}$, $C_{3}$,
$C_{4}$, and $C_{5}$. The error bar on $z$ gives our subjective estimate of
errors of extrapolation. This should be compared with the value $z = 1.623 \pm
0.011$ obtained by Guttmann and Bursill \cite{GUT90}. Because of the larger
fractal dimension of walks, for the same value of $N$, there are significantly
more longer loops generated in $d = 3$ than in $d = 2$. As a result, we see
power-law scaling over roughly $5$ decades of $\ell$ in Fig.~\ref{F:Fl3d}
compared to that of about $4$ decades of $\ell$ in Fig.~\ref{F:Fl2d}. 

\begin{figure}
\centerline{\epsfxsize\hsize\epsfbox{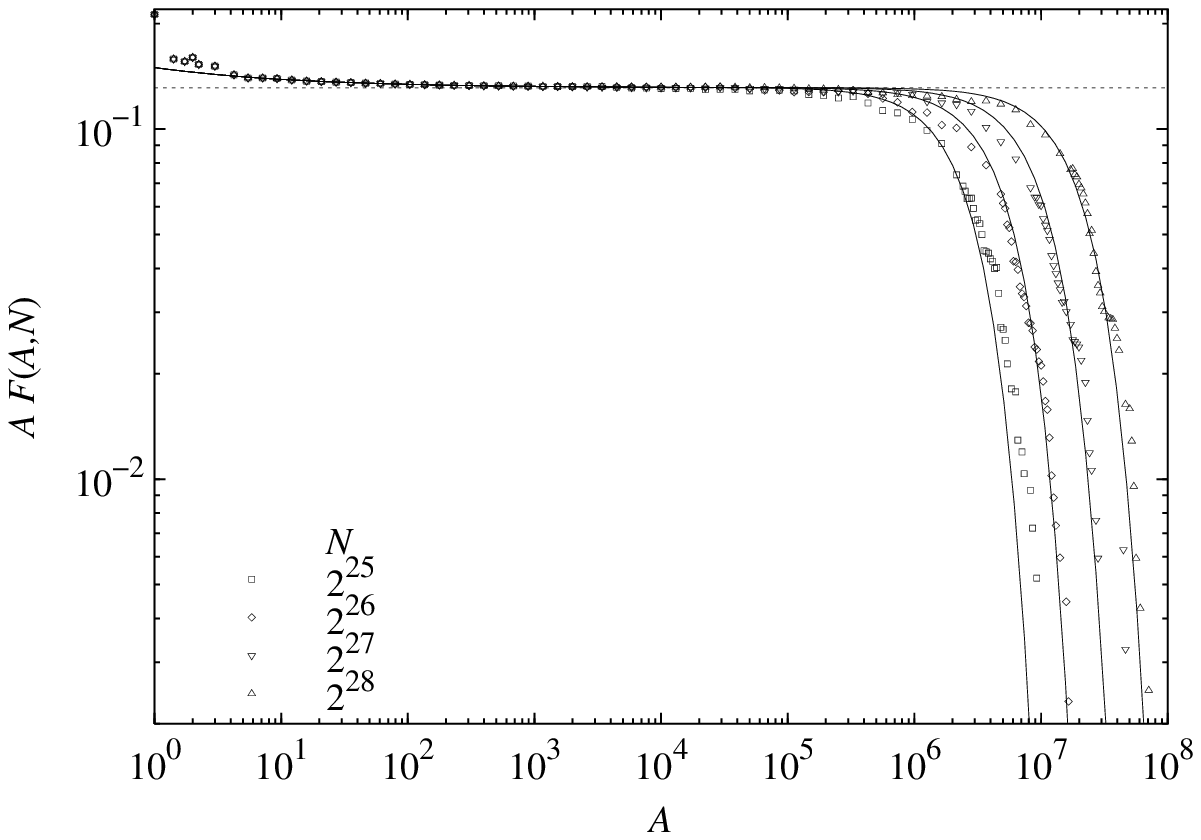}}
\vskip1ex

\caption{Variation of $A F(A,N)$ with $A$ for three-dimen\-sional LERW. Solid
lines represent the best fit of Eq.~(\ref{E:F0}) with parameter values given in
Table~\ref{T:EstPar}.}

\label{F:FA3d}
\end{figure}

In Fig.~\ref{F:FA3d} we show the plot for $F(A,N)$ the cumulative distribution
function for loop area. We have plotted $A F(A,N)$ versus $A$ for different
values of $N$. An unbiased estimate of $C_2$ from the best-fit gives a value
$1.00000 \pm 0.00005$. Thus, we put $C_{2}$ to be exactly $1$ and estimated the
remaining parameters by fitting the scaling form given by Eq.~(\ref{E:F0}) with
$\ell$ substituted by $A$. From the figure it is clearly seen that this form
approximates the entire data very well for $A \gtrsim 10$. The estimated values
of parameters is tabulated in Table~\ref{T:EstPar}. Here also, the exponent in
the correction to scaling term turn out to be different from $1$.

\begin{figure}
\centerline{\epsfxsize\hsize\epsfbox{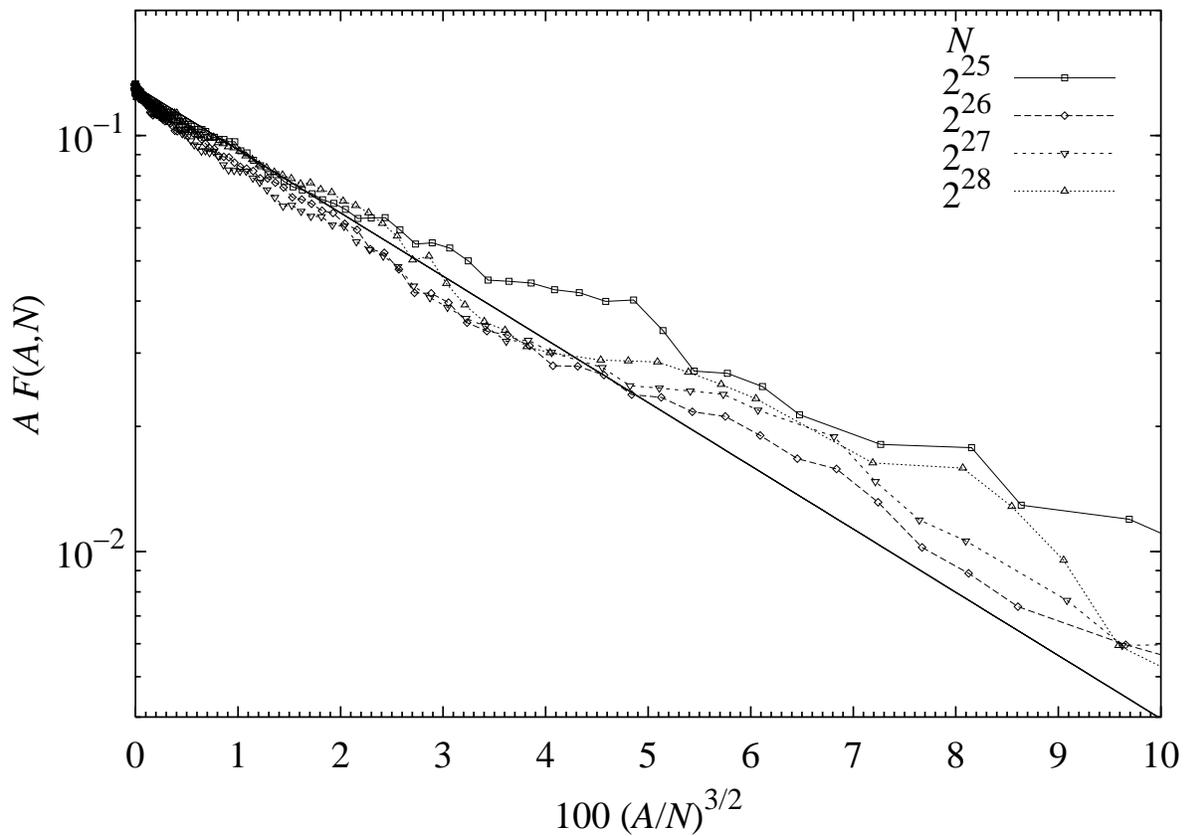}}
\vskip1ex

\caption{Variation of $A F(A,N)$ with $100 (A/N)^{3/2}$ for three\-dimensional
LERW showing the behavior of the exponential correction term. Thick straight
line represents the best fit of Eq.~(\ref{E:F0}) with parameter values given in
Table~\ref{T:EstPar}.}

\label{F:FA3dnl}
\end{figure}

In Fig.~\ref{F:FA3dnl} we have replotted the data of Fig.~\ref{F:FA3d} with $A
F(A,N)$ plotted against $(A/N)^{3/2}$. We see that the curves are approximately
linear for small $A/N$, verifying the theoretical prediction of
Eq.~\ref{E:gsmall}. For larger values of $A/N$, the slope decreases as expected
from Eq.~\ref{E:glarge}.

\begin{figure}
\centerline{\epsfxsize\hsize\epsfbox{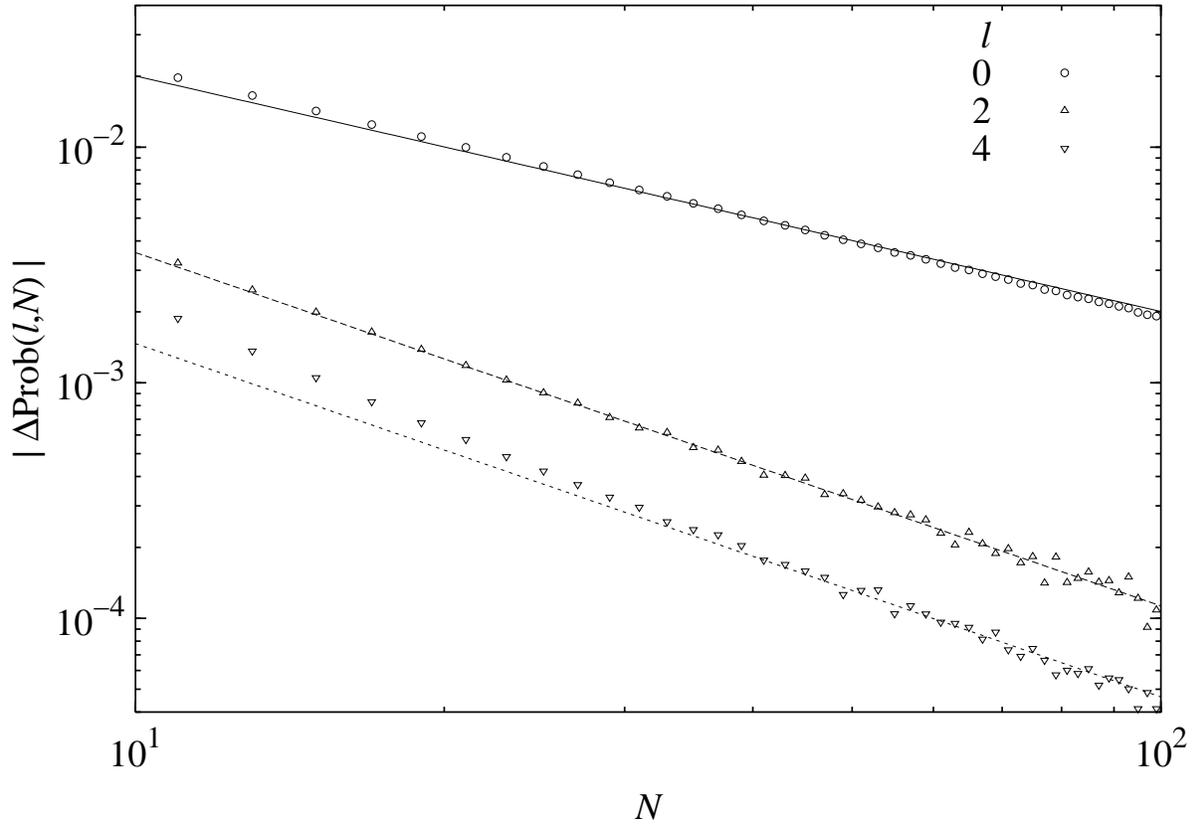}}
\vskip1ex

\caption{Variation of $\Delta \text{Prob}(\ell,N)$ with $N$ for $\ell= 0$, $2$,
and $4$ in three dimensions. Straight lines show the fits using slopes $-1$ for
$\ell = 0$ and $-3/2$ for $\ell = 2$ and $4$.} 

\label{F:dP3d}
\end{figure}

In Fig.~\ref{F:dP3d}, we have plotted the variation of $\Delta
\text{Prob}(\ell,N)$ versus $N$ for $\ell = 0$, $2$, and $4$. The data was
obtained by averaging over $10^{9}$ different realizations of $100$-stepped
walks. For $\text{Prob}(\ell)$, we used the values from the $N = 2^{28}$
simulation. We see good agreement with the predicted $1/N$ variation for $\ell =
0$, and $1/N^{3/2}$ variation for $\ell = 2$ and $4$.

\section{Relation to Exponents of the Sandpile Model} \label{Sec:Sand}

The sandpile model of Bak-Tang-Wiesenfeld is defined as follows \cite{BTW}: We
consider a hypercubical lattice of linear size $L$ in $d$ dimensions. At each
site is a non-negative integer which gives the ``height'' of the pile at that
point. The system is driven by adding a grain of sand at a randomly chosen site,
thereby increasing the height of pile at that site by one. If the height at any
site exceeds $(2d-1)$, it topples, and its height decreases by $2d$, and one
grain is transferred to each of its neighbors. If this makes some other sites
unstable, they are toppled in turn, until all sites are stable, and then a new
grain of sand is added.

Bak {\em et~al.\/} observed that in the steady state of such a pile, adding a
grain gives rise to a sequence of topplings, and the size of such avalanches is
a random variable with a power-law tail. Determining the exact values of the
exponents characterizing these tails has been the main theoretical problem in
the area of self-organized criticality.

The model in $d=1$ is rather trivial, and does not show simple power-law tails
of avalanche sizes, as most avalanches are large \cite{1dBTW}. In $d =2$, unlike
in $d=1$, most avalanches are finite, but they involve multiple topplings of
sites. A theoretical understanding of this case remains incomplete \cite{2dBTW}.
For $d \geq 4$, mean-field description of avalanche propagation is adequate, and
the corresponding exponents are the same as of sizes of clusters in critical
percolation theory \cite{PR99}.

The BTW model for $d=3$ does not suffer from the problems caused by multiple
topplings. It is thus the simplest undirected model for studying self-organized
criticality with nontrivial (non-mean-field) critical behavior. In this case,
multiple topplings at a site occur with very low probability, and the avalanche
clusters are found to be compact, with fractal dimension $3$. Then, simple
scaling arguments \cite{zhang,KTIT00} show that if the probability that there
are exactly $s$ topplings in an avalanche in a system of linear size $L$ is
$\text{Prob}(s\mid L)$, which satisfies the simple finite-size scaling form
\begin{equation} 
  \text{Prob}(s\mid L) \sim s^{-a} h\bigl(s/L^{b}\bigr) 
\end{equation} 
then we must have $a= 4/3$ and $b=3$.

The theoretical assumptions that go into the scaling argument have been checked
extensively in simulations, but a rigorous theoretical proof is not yet
available. Since the number of distinct toppled sites is assumed to be
proportional to the number of topplings, we see that the probability that an
avalanche has $s_{d}$ distinct sites toppled also varies as $s_{d}^{-4/3}$.

As a check on the scaling theory, note that the probability that avalanche
reaches a distance $R$ scales as the probability that number of topplings is
greater than $R^{3}$, hence as $1/R$, which also agrees with the known result
about expected number of topplings at a distance $R$.

The only exponent which this simple argument does not give is the exponent for
the duration of avalanches. But the propagation of avalanches occurs along
spanning trees path in the equivalence between the sandpile model and spanning
trees \cite{DD99}. Hence, the duration $T$ of an avalanche must vary with its
linear extent as $T \sim R^{z}$. And the $z$ is for spanning trees, which is the
same as the $z$ we used for LERWs. The knowledge of $z$, thus, allows us to
estimate the exponent for duration of avalanches: the probability that the
duration of avalanche is greater than $T$ varies as $T^{-y}$, where 
\begin{equation}
  y = 1/z = 0.61795 \pm 0.00015
\end{equation}

\section{Concluding Remarks} \label{Sec:Conc}

We have already noted that the LERW problem is very suited for numerical
studies. In the two-dimensional case, we have collected data for over $10^3$
realizations of walks with $N$ up to $2^{29}$. Thus the numerically determined
loop-size distribution is an average over more than $8 \times10^{10}$ loops
(only about $31.25\%$ of the steps taken form non-trivial loops on square
lattice). For the three-dimensional case, the corresponding number is $2.8
\times 10^{10}$ loops (only about $21.17\%$ steps form non-trivial loops on
cubic lattice). The quantity which corresponds closest to loop-erasures are
avalanches in the sandpile model (more correctly, subavalanches)
\cite{shcherbakov}. Clearly, simulation of the Abelian sandpile model with equal
number of avalanches is not possible with available computing machines.

Secondly, our simulations are done on an effectively infinite lattice, and there
are no complicated boundary effects to complicate the analysis of data.
Corrections due to finite size of system show up only in the finiteness of the
number of steps $N$ of the random walk. This seems to be well described by
simple finite-size scaling theory. If we wanted to determine the exponent $z$
using the sandpile model, or the spanning trees, the largest system sizes
accessible would be much smaller.

The dimension-independence of the exponent characterizing the distribution of
areas of erased loops for $2 \le d \le 4$ is rather unexpected. The exponent
does depend on dimension for $d > 4$. We have been unable to find a more
transparent proof of this result.

We would like to thank S. N. Majumdar for his critical reading of an earlier
version of this paper.

\mediumtext

\begin{table*}

\caption{Estimated values of various parameters corresponding to
Eq.~(\ref{E:F0}) for variation of loop perimeter and area for two- and
three-dimensional LERWs. The values without error bars are fixed during
estimation of other parameters.}

\vskip2ex

\label{T:EstPar}

\begin{tabular}{lcccccc}
    &        & $C_{1}$
    	       & $C_{2}$
	         & $C_{3}$
		   & $C_{4}$
		     & $C_{5}$  \\
\hline\hline                                                                                                
2-D & $\ell$ & $0.3533\pm0.0004$
	       & $1.5997\pm0.0005$
	         & $1.1\pm0.1$
		     & $1.58\pm0.03$
		       & $1.000\pm0.007$ \\
    & $\ell$ & $0.35385\pm0.00025$
	       & $8/5$
	         & $1.1\pm0.1$
		     & $1.56\pm0.03$
		       & $1$ \\
    & $A$    & $0.127316\pm0.000015$
	       & $1$
		 & $9.8\pm0.7$
		     & $0.494\pm0.003$
		       & $1$ \\
\hline                                                                                                      
3-D & $\ell$ & $0.1527\pm0.0003$
               & $1.2359\pm0.0003$
	         & $2.8\pm0.3$
		     & $1.69\pm0.02$
		       & $0.86\pm0.05$ \\
    & $A$    & $0.1312\pm0.0002$
               & $1$
	         & $35\pm7$
		     & $0.142\pm0.005$
		       & $0.394\pm0.015$ \\
\end{tabular}
\end{table*}

\narrowtext

\begin{references}
\bibitem{footnote} On a linear chain, only possible loops are of length zero or
two. Longer loops do occur on more complicated one-dimensional graphs, such as
ladders.
\bibitem{note1} To be precise, the function plotted is an average of
$F(\ell,N')$ with $N'$ lying between $N/2$ and $N$.
\end{references}
\end{document}